\documentclass[trackchanges,twocolumn]{aastex701}%

\usepackage{amssymb}
\usepackage{xcolor}

\newcommand{\cmark}{\textcolor{green!60!black}{$\checkmark$}}
\newcommand{\xmark}{\textcolor{red!70!black}{$\times$}}

\newcommand{\nrc}{NRC Herzberg Astronomy and Astrophysics,
5071 West Saanich Road,
Victoria, BC, V9E 2E7, Canada}
\newcommand{\ucsd}{Department of Astronomy \& Astrophysics,  University of California, San Diego, La Jolla, CA 92093, USA}
\newcommand{\nmsu}{Department of Astronomy, New Mexico State University, 1320 Frenger Mall, Las Cruces, NM 88003, USA}
\newcommand{\uvic}{Department of Physics and Astronomy, University of Victoria,
3800 Finnerty Road, Elliott Building,
Victoria, BC, V8P 5C2, Canada}
\newcommand{\mac}{Department of Physics \& Astronomy, McMaster University, 1280 Main Street West, Hamilton, ON, L8S 4L8, Canada}

\newcommand{\ucsc}{Department of Astronomy \& Astrophysics, University of California, Santa Cruz, CA 95064, USA}
\newcommand{\ucsb}{Department of Physics, University of California, Santa Barbara, CA 93106, USA}
\newcommand{\ucla}{Department of Earth, Planetary, and Space Sciences, University of California, Los Angeles, CA 90095, USA}
\begin{document}

\title{Dynamical Mass Constraints on Transition Disk Perturbers with the G23H Catalog}

\author[0000-0001-9582-4261]{Dori Blakely}
\affiliation{\uvic}
\affiliation{\nrc}
\email[show]{blakelyd@uvic.ca}

\author[0000-0001-5684-4593]{William Thompson}
\affiliation{\nrc}
\email{william.thompson@nrc-cnrc.gc.ca}  

\author[0000-0002-6773-459X]{Doug Johnstone}
\affiliation{\nrc}
\affiliation{\uvic} 
\email{Douglas.Johnstone@nrc-cnrc.gc.ca}

\author[0000-0003-3430-3889]{Jessica Speedie}
\altaffiliation{Heising-Simons Foundation 51 Pegasi b Fellow}
\affiliation{\uvic}
\affiliation{Department of Earth, Atmospheric, and Planetary Sciences, Massachusetts Institute of Technology, Cambridge, MA 02139, USA}
\email{jspeedie@mit.edu}

\author[0000-0002-6618-1137]{Jerry W. Xuan}
\altaffiliation{Heising-Simons Foundation 51 Pegasi b Fellow}
\affiliation{\ucla}
\email{jerryxuan@g.ucla.edu}  

\author[0000-0002-9632-1436]{Simon Blouin}
\affiliation{\uvic}
\email{sblouin@uvic.ca}

\author[0000-0002-2696-2406]{Jingwen Zhang}
\affiliation{\ucsb}
\email{jwzhang@ucsb.edu}

\author[0000-0003-2233-4821]{Jean-Baptiste Ruffio}
\affiliation{\ucsd} 
\email{jruffio@ucsd.edu}

\author[0000-0001-6975-9056]{Eric Nielsen}
\affiliation{\nmsu}
\email{nielsen@nmsu.edu}

\author[0000-0003-2649-2288]{Brendan P. Bowler}
\affiliation{\ucsb}
\email{bpbowler@ucsb.edu}

\author[0000-0003-4557-414X]{Kyle Franson}
\altaffiliation{NHFP Sagan Fellow}
\affiliation{\ucsc}
\email{kfranson@ucsc.edu}

\author[0009-0008-9687-1877]{William Roberson}
\affiliation{\nmsu}  
\email{wcr@nmsu.edu}

\author[0000-0001-5383-9393]{Ryan Cloutier}
\affiliation{\mac}
\email{ryan.cloutier@mcmaster.ca}

\author[0009-0007-6766-2040]{Andre Fogal}
\affiliation{\uvic}
\email{afogal@uvic.ca}

\author[0009-0009-2223-2404]{Kaitlyn Hessel}
\affiliation{\uvic}
\email{khessel@uvic.ca}

\author[0000-0002-4164-4182]{Christian Marois}
\affiliation{\nrc}
\affiliation{\uvic}
\email{Christian.Marois@nrc-cnrc.gc.ca}

\author[0009-0008-1229-3230]{Alexandra Rochon}
\affiliation{\mac}
\email{rochoa3@mcmaster.ca}

\begin{abstract}
{We present dynamical mass constraints on perturbers in 11 transition disk systems using a novel combination of calibrated Hipparcos and Gaia absolute astrometry data. Out of the sample of 11, we find support for companions in four systems, with significant detections in two. These systems are: HD 142527, where we clearly detect the known low-mass stellar companion HD 142527 B and MWC 758, where we detect a likely sub-stellar companion. We also find moderate evidence of companions to AB Aur and UX Tau A. For the seven systems with non-detections, we find no evidence for companions more massive than $\sim$12 $M_{\mathrm{Jup}}$ with a semi-major axis greater than 3 au for both HD 100546 and HD 100453, nor for companions more massive than $\sim$3 $M_{\mathrm{Jup}}$ with a semi-major axis greater than 2 au for TW Hya. We also find no evidence for stellar mass companions with semi-major axes between $\sim$4 and $\sim$25 au for HD 34282, HD 97048, CQ Tau and RY Lup. In addition to our fiducial model, we perform cross validation between astrometry sources. By comparing results across models, we find tentative evidence of a short timescale excess astrometric noise that may impact some protoplanetary disk systems. We conclude with predictions for the prospects of making dynamical mass constraints on protoplanets in protoplanetary disk systems with Gaia data release 4 using detailed simulations of Gaia DR4 data of PDS 70 and WISPIT 2.}

\end{abstract}



\section{Introduction}
\label{sec:introduction}



Absolute astrometry from Hipparcos \citep{1997A&A...323L..49P} and Gaia \citep{2023A&A...674A...1G} has been successfully used to discover and constrain the mass of {fully formed} sub-stellar companions around many young stars \citep[e.g.,][]{2018ApJS..239...31B,2019A&A...623A..72K,2021ApJS..254...42B,2022A&A...657A...7K,2023A&A...672A..93M,2023A&A...672A..94D,2023ApJ...950L..19F,2023Sci...380..198C,2026AJ....171....5C}. 
Thus far, these techniques have not widely been used to study {systems with} protoplanetary disks, partly due to concern about astrometric bias due to disk material inhomogeneously affecting the photocenter \citep{2022RNAAS...6...18F,2025arXiv250110488L}. 
Nonetheless, absolute astrometry could be an important technique for detecting planets in protoplanetary disks because such companions may be hidden from direct imaging due to significant local extinction \citep{2025AJ....170..317C}, and hidden from radial velocity (RV) and transit detection due to limitations such as stellar activity and disk contamination, unless there is a significant disk-orbit misalignment \citep[e.g.,][]{2024Natur.635..574B}.

In spite of concerns about astrometric bias due to disk material, two recent papers have shown that absolute astrometry can be used to discover companions in protoplanetary disk systems. These works use the difference between the Gaia DR2 and DR3 catalog (hereafter GDR2 and GDR3) proper motions to infer the presence of a $\sim$10 $M_{\mathrm{Jup}}$ planet around PDS 66 \citep{2025NatAs...9.1176R} and to detect 31 companion candidates (including signals compatible with several previously known companions) out of a sample of 98 transition disks \citep{2025arXiv251200157V}. These Gaia-only studies use a method that has primarily been used for---and validated on---identifying binary stars \citep{2022MNRAS.513.2437P,2022MNRAS.516.3661A}, and may be limited by known systematic biases that, while small for binary stars, could be significant for sub-stellar companions. 

Our concern arises because the methodology used in these two papers is based on the uncalibrated GDR2 and GDR3 catalog proper motions values, and their associated uncertainties. The GDR2 and GDR3 catalogs have different reference frames, prominently due to a $\sim$0.1 mas yr$^{-1}$ rotation---significant at the 2$\sigma$ level---relative to background quasars for GDR2 \citep{2020A&A...633A...1L} that has been corrected to a level of several $\mu$as yr$^{-1}$ in GDR3 \citep{2022A&A...667A.148G}. Additionally, the raw GDR2 and GDR3 proper motion uncertainties have been found to be underestimated by a multiplicative factor of $\sim$1.74 for GDR2 \citep{2018ApJS..239...31B} and $\sim$1.37 for GDR3 \citep{2021ApJS..254...42B}. Therefore, 
a methodology that incorporates these details is crucial for robust detections and reliable mass constraints, particularly when pushing to the planetary mass regime. 

In this paper, we search for and constrain the astrometric signature of companions in a sample of transition disks using the Calibrated Gaia DR2, DR3, and Hipparcos catalog (G23H) \citep{thompson2026detectingcharacterizingcompanionscalibrated}. The G23H catalog includes a collection of astrometric constraints that include { Hipparcos intermediate astrometry data (IAD) \citep{2007A&A...474..653V,2020AJ....159...71N}}; calibrated Hipparcos proper motions; calibrated GDR3-Hipparcos position differences and GDR3 proper motions \citep[the HGCA,][]{2021ApJS..254...42B}; a calibrated model of the GDR3 astrometric excess noise, the unbiased estimator of variance (UEVA) \citep[][]{2025A&A...702A..76K}; along with GDR2 proper motions and positions that are calibrated to the GDR3 reference frame \citep{thompson2026detectingcharacterizingcompanionscalibrated}. 

{We apply the G23H catalog to a sample of transition disks that are in the Hipparcos catalog as well as the Gaia catalogs. We limit our sample to systems that are included in the Hipparcos catalog because this allows for us to preliminarily empirically examine the effect of disk material on the Gaia absolute astrometry by comparing the short-timescale information provided by Gaia to the long-timescale information provided by the difference between the Hipparcos and Gaia measurements. Due to this restriction, we note that our sample is a subset of the transition disks looked at by \citet{2025arXiv251200157V} using Gaia data alone.}



 Additionally, we simulate Gaia data release 4 (GDR4) data of known protoplanet hosts PDS 70 and WISPIT 2 \citep{2018A&A...617A..44K,2019NatAs...3..749H,2025ApJ...990L...8V,2025ApJ...990L...9C}, using realistic GDR4 noise estimates calculated from GDR3 data with the UEVA \citep[][]{2025A&A...702A..76K}. With the results of the GDR4 simulations, we explore the expected dynamical mass constraints on protoplanets in protoplanetary disks that will be possible with GDR4, and compare these limits to what is currently possible using the G23H catalog. Based on these results, 
 and lingering concerns about contamination due to disk material, we suggest that comprehensive astrometry-based protoplanet demographics studies of protoplanetary disk-hosting stars will have to wait until the upcoming release of GDR4.

This paper is structured as follows: \S\,\ref{sec:methods} outlines the methods and the models; \S\,\ref{sec:results} outlines the results; \S\,\ref{sec:discussion} explores the implications of our results and discusses expectations for GDR4. We close with a summary in \S\,\ref{sec:conclusions}.

\section{Methods} \label{sec:methods}

\begin{table*}[t]
\centering
\caption{G23H catalog parameters and constraints used by each model.}
\label{tab:model_components}
\begin{tabular}{lccccccccccc}
\hline\hline
Model
& HIP
& HG
& DR2
& DR32
& DR3
& UEVA
& Gaia jitter
& Co-planar
& DR3 deflation
& RV variability
& Figures\\ \hline
jitter (fiducial)
& \cmark & \cmark & \cmark & \cmark & \cmark
& \xmark & \cmark & \xmark & \xmark & \xmark & \ref{fig:mass-a-nogeom-jitter}, \ref{fig:jitters} \\

co-planar, jitter
& \cmark & \cmark & \cmark & \cmark & \cmark
& \xmark & \cmark & \cmark & \xmark & \xmark & \ref{fig:disk-clear-detections}, \ref{fig:full-pos-preds}, \ref{fig:mass-a-geom-jitter} \\

no jitter
& \cmark & \cmark & \cmark & \cmark & \cmark
& \xmark & \xmark & \xmark & \xmark & \xmark & \ref{fig:mass-a-nojitter} \\

UEVA
& \cmark & \cmark & \cmark & \cmark & \cmark
& \cmark & \xmark & \xmark & \xmark & \xmark & \ref{fig:mass-a-UEVA} \\ \hline
\end{tabular}
\end{table*}

Using the G23H catalog \citep{thompson2026detectingcharacterizingcompanionscalibrated}, and the orbit modelling code \texttt{Octofitter}\footnote{https://sefffal.github.io/Octofitter.jl/} \citep{2023AJ....166..164T} we explore the dynamical mass constraints that we can place on companions in 11 transition disk systems: AB Aur, CQ Tau, HD 34282, HD 97048, HD 100453, HD 100546, HD 142527, MWC 758, RY Lup, TW Hya, and UX Tau A. 
We sample from the posterior of each model using non-reversible parallel tempering \citep{https://doi.org/10.1111/rssb.12464}, as is implemented in \texttt{Pigeons.jl} \citep{surjanovic2023pigeons}. We run \texttt{Pigeons} for 12 rounds (4096 final samples after ``burn-in") with {16} chains using the slice sampler explorer, ensuring convergence for all models. 

We run four distinct models using different constraints and/or subsets of the G23H catalog measurements. We do this primarily for cross-validation and to explore whether we observe any potential contamination due to disk material on the calibrated proper motions and on the UEVA \citep[][]{2025A&A...702A..76K}, which could manifest as significant differences in the constraints found for each model. {Additionally, for completeness, in Appendix \ref{sec:hgca_model}, we also present results from a model consisting of only the HGCA and the Hipparcos IAD, that we call the \textit{HGCA model}.} 

Our fiducial model, the \textit{jitter model} consists of the {calibrated HGCA Hipparcos proper motion, the Hipparcos IAD \citep{2007A&A...474..653V,2020AJ....159...71N}}, the G23H calibrated GDR2 proper motion, G23H calibrated GDR3-GDR2 position difference (GDR32), {the HGCA GDR3-Hipparcos position difference and the GDR3 proper motion \citep{2021ApJS..254...42B}, but does not include the G23H UEVA constraint, the \texttt{paired} radial velocity variability constraint \citep{2025ApJ...992..131C}}, or the GDR3 uncertainty deflation calculated using the UEVA, due to concerns that these components may be contaminated by bright disk material. Additionally, we include a fitted error inflation, jitter term, applied to all of the Gaia proper motions with the functional form given by $\sigma_{\mu} = \sigma_{\mu,catalog} \sqrt{\left(1 + \sigma_{\mu,jitter}^2 \right)}$. Note that we assume that this factor is negligible compared to the Hipparcos uncertainties and thus, we do not apply this term to the Hipparcos-Gaia long-term proper motion uncertainties. The other three models that we explore are: the \textit{co-planar, jitter model}, which incorporates a constraint on the geometry of the companion to be near co-planar with the disk (including the known disk orbital directions), enforced using a Gaussian prior on the mutual inclination between the orbital plane and the disk plane, with a standard deviation of 10 degrees, following \citet{2020AJ....159..263W,2021AJ....161..148W}; a model without the co-planar constraint and without the error inflation, the \textit{no jitter model}; and a model with the UEVA, without the error inflation and without the co-planar constraint, the \textit{UEVA model}. We summarize which G23H parameters and constraints each model uses in Table \ref{tab:model_components}.

The priors that we adopt on the orbital elements are outlined in Table \ref{tab:priors}. The adopted stellar masses, disk geometries (including the orbital direction of the disk) and cavity sizes are given in Table \ref{tab:geometry}. We note that for sampling (and the presented posteriors) we are assuming dark companions to avoid making assumptions about the flux ratio of any companions. This means that we would slightly underestimate the mass of unresolved companions, though at a level that is negligible except for near equal mass companions. Additionally, we note that near-equal brightness companions can produce identical astrometric signatures to low mass dark companions, which will be missed by our models. See Section 8.4 from \cite{2025A&A...702A..76K} for an example showing the effect of modelling dark versus luminous companions. 

We estimate the significance of detections using Bayes factors \citep{Kass01061995}, or equivalently odds ratios. We calculate Bayes factors for each model by using 
a binary indicator variable $z \sim$ Bernoulli(0.5), {included in each model}, that is multiplied by the companion mass, such that $z = 0$ corresponds to the zero companion model and $z = 1$ corresponds to the single companion model. 
We calculate Bayes factors, assuming that each model (companion or no companion) is equally likely \emph{a priori}, from the posterior mean of $z$, $\overline{z}$, using BF $= \overline{z}/(1-\overline{z})$.

In addition to the 11 transitions disks that are included in the Hipparcos catalog, our main sample, we also look at subsets of the G23H Gaia-only constraints for the stars PDS 70 and WISPIT 2 that are not included in the Hipparcos catalog. We compare the GDR2 and GDR3 only predictions to what we expect to be possible with the epoch astrometry released with GDR4 by running detailed simulations of GDR4 data for both targets. We outline our GDR4 simulation procedure and discuss these results in Section \ref{sec:discussion}.

\begin{table}[h!]
\centering
\renewcommand{\arraystretch}{1.2}
\begin{tabular}{ll}
\hline
\textbf{Prior} & \textbf{Units} \\ \hline
$ a \sim \mathrm{LogUniform}(0.1,\;150)$ & au \\
$ e \sim \mathrm{Uniform}(0,\;0.99)$ & --- \\
$ i^a \sim \mathrm{Uniform}(0,\;2\pi)$ & rad \\
$ \Omega^a \sim \mathrm{Uniform}(0,\;2\pi)$ & rad \\
$ \omega \sim \mathrm{Uniform}(0,\;2\pi)$ & rad \\
$ \theta \sim \mathrm{Uniform}(0,\;2\pi)$ & rad \\
$ M_{\mathrm{pri}} \sim \mathcal{N}\!\bigl(M_{\mathrm{pri,val}},\;0.1 \cdot\,M_{\mathrm{pri,val}}\bigr)$ & $M_\odot$ \\
$ M_{b} \sim \mathrm{LogUniform}\!\left(0.1,\;M_{\mathrm{pri,val}}\right)$ & $M_{\mathrm{Jup}}$ \\
$ \varpi \sim \mathcal{N}\!\bigl(\varpi_{\mathrm{DR3}},\;\sigma_{\varpi,\mathrm{DR3}}\bigr)$ & mas \\
$ \mu_{\alpha^\ast} \sim \mathrm{Uniform}\!\bigl(\mu_{\alpha^\ast,\mathrm{DR3}}-1,\;\mu_{\alpha^\ast,\mathrm{DR3}}+1\bigr)$ & mas\,yr$^{-1}$ \\
$ \mu_{\delta} \sim \mathrm{Uniform}\!\bigl(\mu_{\delta,\mathrm{DR3}}-1,\;\mu_{\delta,\mathrm{DR3}}+1\bigr)$ & mas\,yr$^{-1}$ \\
$ \sigma_{\mu,jitter} \sim \mathrm{LogUniform}\!\left(10^{-9},\;10\right)$ & --- \\
\hline

\end{tabular}
\caption{Priors on the model parameters. See \citet{thompson2026detectingcharacterizingcompanionscalibrated} for details on the additional nuisance parameters. \\
$^a$ A Gaussian prior with a standard deviation of 10 degrees on the mutual inclination between the orbital plane of the companion and the orbital plane of the disk was additionally included for the \textit{co-planar, jitter model}.}
\label{tab:priors}
\end{table}

\begin{table*}[ht]
\centering
\renewcommand{\arraystretch}{1.2}
\begin{tabular}{lcccccc|c}
\hline
{Disk} & Distance$^{*}$ (pc) & {Mass ($M_\odot$)} & {Inc ($^\circ$)} & {PA ($^\circ$)} & {$R_{\rm cav,mm}^{**}$ (au)} & {References} &  {BF$^{***}$} \\
\hline
HD 142527 & $159.3 \pm 0.7$ & 2.0 & 141.8 & 162.7 & 180 & ${2}$, ${5}$, ${6}$  &   $>$$4096$\\
MWC 758 & $155.8 \pm 0.8$ & 1.40 & 160.6 & 240.3 & 50 & ${1}$, ${2}$  &   90.0\\
AB Aur & $155.9 \pm 0.9$ & 2.23 & 23.2 & 236.7 & 170 & ${2}$, ${8}$, ${9}$  &   7.6\\
UX Tau A & $142.3 \pm 0.7$ & 1.4 & 40 & 347 & 31 & ${3}$, ${13}$  &   4.1\\
\hline
CQ Tau & $149.4 \pm 1.3$ & 1.40 & 36.3 & 235.1 & 50 & ${1}$, ${2}$  &    2.0\\
HD 97048 & $184.4 \pm 0.9$ & 2.17 & 41 & 4 & 63 & ${3}$, ${12}$  &   1.3\\
RY Lup & $153.5 \pm 1.4$ & 1.4 & 67 & 289 & 69 & ${3}$, ${4}$, ${14}$ &   1.2 \\
HD 100546 & $108.1 \pm 0.5$ & 2.2 & 41.7 & 323.0 & 27 & ${3}$, ${7}$ &   0.8 \\
HD 34282 & $308.6 \pm 2.3$ & 1.62 & 121.7 & 117.4 & 87 & ${1}$, ${3}$  &  0.8\\
HD 100453 & $103.8 \pm 0.2$ & 1.47 & 30 & 329 & 30 & ${3}$, ${4}$, ${11}$  &  0.6\\
TW Hya & $60.1 \pm 0.1$ & 0.81 & 173 & 155 & 12 & ${2}$, ${3}$, ${4}$, ${10}$  &  0.5\\
\hline

\end{tabular}
\caption{Disk parameters compiled from the literature, and our Bayes factor (BF) or odds ratio of companion presence over no companion.  We use a 10\% uncertainty on the stellar mass values. $^{*}$The distance values and uncertainties are the marginal posterior mean and standard deviation from the \textit{jitter model}. $^{**}$When available we use the inner-most $R_{\mathrm{dust}}$ value from \citet{2021AJ....161...33V}. Otherwise, we use $R_{\mathrm{cav}}$ from \citet{2020ApJ...892..111F}. $^{***}$The Bayes factor is calculated comparing a single companion model to a zero companion model for the \textit{jitter model}. \newline References: $^{1}$ \cite{2025ApJ...984L...8I}, $^{2}$\cite{2021AJ....161...33V}, $^{3}$\cite{2020ApJ...892..111F}, $^{4}$\cite{2023A&A...670A.154W}, $^{5}$\cite{2014ApJ...790...21M}, $^{6}$\cite{2022A&A...658A.183B}, $^{7}$\cite{2022ApJ...933L...4C}, $^{8}$\cite{2017ApJ...840...32T}, $^{9}$\cite{2024Natur.633...58S}, $^{10}$\cite{2019ApJ...884L..56T}, $^{11}$\cite{2018ApJ...854..130W}, $^{12}$\cite{2016ApJ...831..200W}, $^{13}$\cite{2020A&A...639L...1M}, and  $^{14}$\cite{2022A&A...668A..25V}.}

\label{tab:geometry}
\end{table*}

\begin{figure*}
\centering
\includegraphics[width=0.82\textwidth]{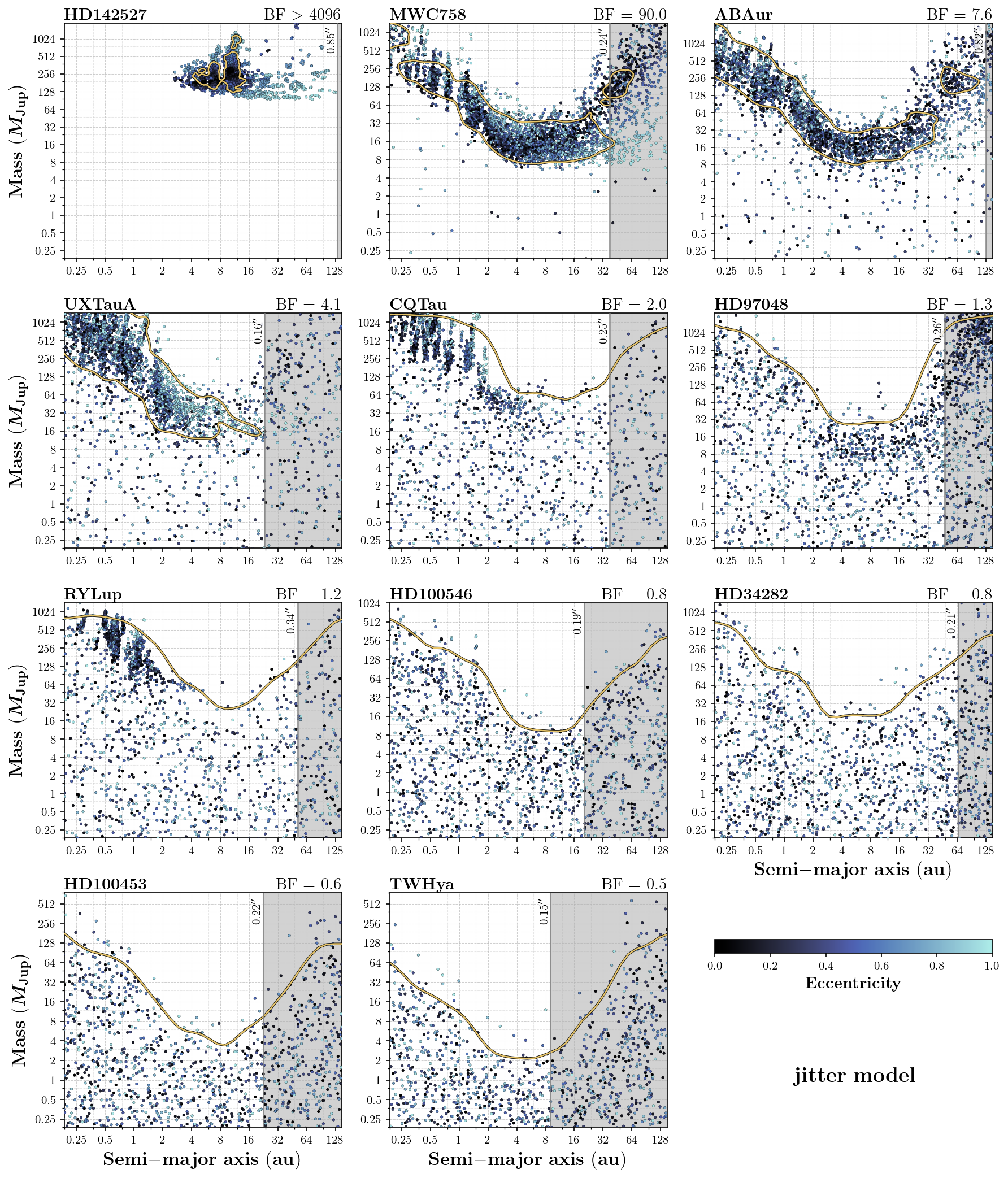}
\caption{Companion mass and semi-major axis marginal posterior for the (fiducial) \textit{jitter model}. Panels are ordered by decreasing Bayes factor (BF). The Bayes factor can can be interpreted as the odds ratio in favour of a single companion model over a zero companion model, assuming both are equally likely \emph{a priori}. For stars with BF $>$ 3, the gold contour represents the 1$\sigma$ posterior density. For stars with BF $<$ 3, the gold curve represents the 95th percentile upper limit. The shaded region shows 0.75$\times R_{cav,mm}$, an empirical estimate of the gas cavity size \citep{2025A&A...698A.102R}.} 
\label{fig:mass-a-nogeom-jitter}
\end{figure*}

\section{Results} \label{sec:results}

We present the mass versus semi-major axis posterior samples from our fiducial model, the \textit{jitter model}, in Figure \ref{fig:mass-a-nogeom-jitter}. With this model, we see some evidence of a companion  (BF $>$ 3) in {four} out of the eleven systems, with {two} of the {four} showing significant evidence (BF $>$ 20). Bayes factors comparing the single companion \textit{jitter model} to the zero companion \textit{jitter model} are listed in Table \ref{tab:geometry} and are displayed in Figure \ref{fig:mass-a-nogeom-jitter}. Bayes factors for the other models are displayed in Figures \ref{fig:mass-a-geom-jitter}, \ref{fig:mass-a-nojitter} and \ref{fig:mass-a-UEVA}, in the Appendix. These Bayes factors can be interpreted as the odds ratio in favour of the single companion model over the zero companion model, under the assumption that each model is equally likely \emph{a priori}. {We note that if we were to only use the Hipparcos IAD and the HGCA, we would only find evidence of companions in two out of the 11 systems, as shown in Appendix \ref{sec:hgca_model}. This result highlights the constraining power of the combination of even individual long-baseline astrometric measurements---the Hipparcos-Gaia long term proper motion, with short-baseline astrometric measurements---in particular, the combination of the GDR3 proper motion, and the calibrated GDR2 proper motion, with a model of the correlation between the GDR2 and GDR3 proper motions \citep{2025AJ....170..301T,thompson2026detectingcharacterizingcompanionscalibrated}. This result is a small preview of the constraining power that will be possible with the combination of GDR4 epoch astrometry and Hipparcos astrometry. }

We discuss in detail all 11 sources, their candidate companions and our analysis in Section \ref{sec:comparisonto_other}. The {two} examples where a significant astrometric signal (BF $>$ 20) is observed are: a clear detection of the known low-mass stellar companion HD 142527 B \citep[e.g.,][]{2012ApJ...753L..38B} 
and a newly detected (likely) sub-stellar companion to MWC 758. The {two} other systems with BF $>$ 3 are {AB Aur} and UX Tau A which have moderate evidence for a companion. For all of the sources with BF $>$ 3 the gold contours in Figure \ref{fig:mass-a-nogeom-jitter} denote the 1$\sigma$ posterior density, which are calculated by smoothing the posterior sample distribution (in log-space) using the \texttt{scipy.stats.gaussian\_kde} function \citep{2020SciPy-NMeth} with a bandwidth factor of 0.1, and estimating what we call the 1$\sigma$ posterior density as the 68\% highest posterior density (HPD) region of this distribution. 
For the {seven} systems with non-detections, {CQ Tau}, {HD100546}, {HD 97048}, RY Lup, HD 100453, HD 34282 and TW Hya (BF $<$ 3), the gold curves denote a 95th percentile upper limit. The same figures for the three additional models are shown in Appendix \ref{sec:mass-a-post}. We discuss differences between the models in Section \ref{sec:comparison}.




\section{Discussion} \label{sec:discussion}


\begin{figure*}
\includegraphics[width=1\textwidth]{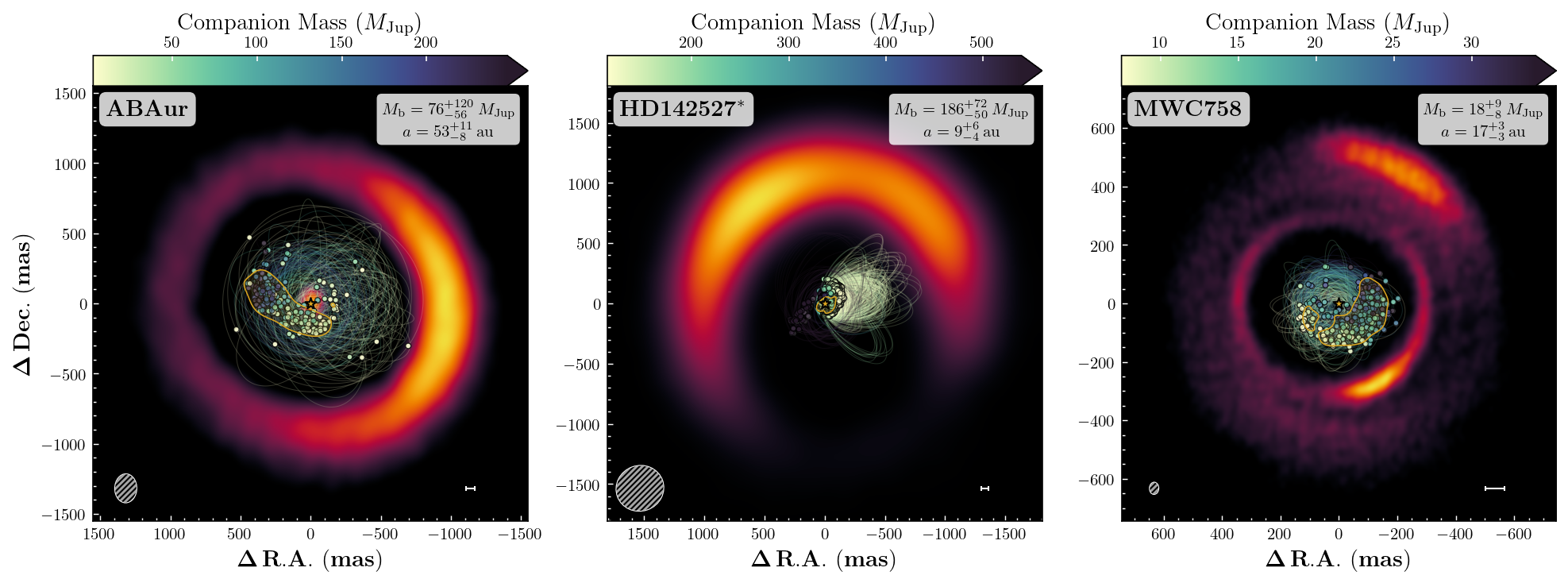}
\caption{Posterior samples from the \textit{co-planar, jitter model} consistent with having cleared the observed cavity (see Section \ref{sec:cav_clear} for details). The dots denote the position of the companion on 1 March 2026 for each individual posterior sample, with the gold contours denoting the 1$\sigma$ posterior density. The dots are coloured by mass, along with their corresponding orbits, with a colour scale that is clipped to the 5th and 95th percentile of the mass distribution for visual clarity. The background images show ALMA continuum images of each disk from \citet{2025ApJ...981L..30S}, 
\citet{2020ApJ...905...89Y} and 
\citet{2018ApJ...860..124D}, with their respective beam sizes shown in the bottom left corner and a 10 au scale bar shown in the bottom right corner. The ALMA images are centered arbitrarily. In the top right corner of each panel we show the median plus/minus the 84th/16th percentiles of the mass and semi-major axis distributions for all of the plotted samples.
\newline
$^*$ We note that we plot all of the posterior samples for HD 142527, because the orbit of HD 142527 B is too compact to explain the size of the observed cavity \citep{2024A&A...683A...6N}.}
\label{fig:disk-clear-detections}
\end{figure*}










\subsection{Comparison between models}
\label{sec:comparison}

\begin{figure}
\includegraphics[width=1\linewidth]{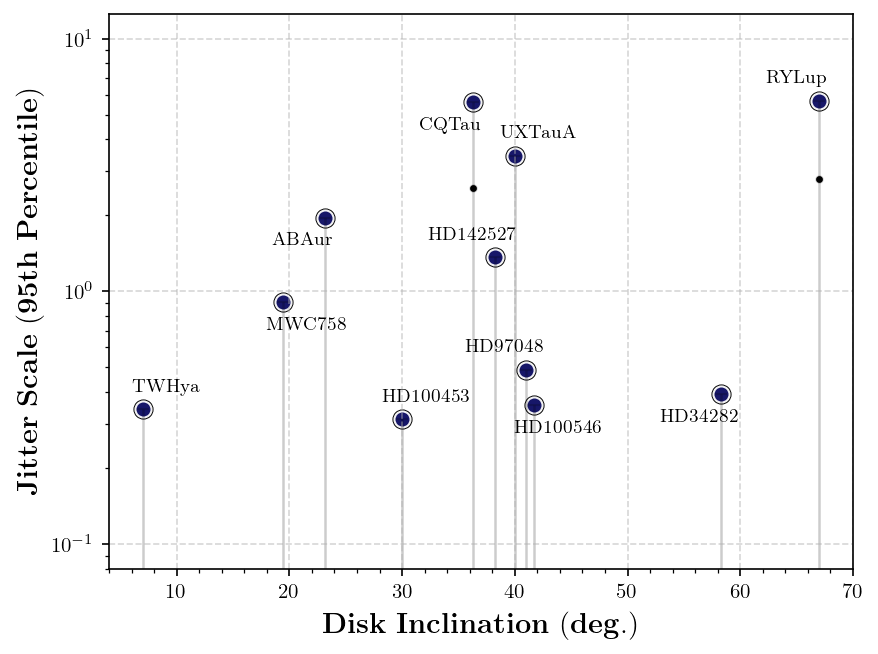}
\caption{We plot the 95th percentile upper limit of the Gaia jitter parameter, $\sigma_{\mu,jitter}$ as a function of the outer disk inclination, shown by the blue filled circles. We also show the median value and the the 5th percentile, where visible, denoted by the black dot, and the error bars. The lack of clear correlation with inclination indicates that this additional noise is independent of the orientation of the outer disk and thus is unlikely to be related to the outer disk. Alternative possible sources of excess astrometric ``noise" could be due to multiple massive companions producing a signal not captured by our single companion models, or possibly due to the signal from a variable, bright inner disk.}
\label{fig:jitters}
\end{figure}





The \textit{co-planar, jitter model} (Figure \ref{fig:mass-a-geom-jitter}, in the Appendix) provides similar but slightly more constrained mass -- semi-major axis predictions compared with the \textit{jitter model} (Figure \ref{fig:mass-a-nogeom-jitter}). The \textit{jitter model} does provide the most general constraints, but due to the limited astrometric information available, it is not possible to significantly localize the position of a companion without additional constraints. However, protoplanets are expected to be near co-planar with their protoplanatary disk, which is the assumption we make with the \textit{co-planar, jitter model}. By applying this additional geometry constraint in the form of a Gaussian prior on the companion's orbital geometry (see Section \ref{sec:methods}), we are able to constrain the position of a perturbing companion, even well past the 2017 end of the GDR3 data range. We demonstrate this in Figure \ref{fig:disk-clear-detections} with predictions for 1 March 2026 (and Figure \ref{fig:full-pos-preds}, in the Appendix, with a comparison between 2015 and 2026 predictions), which we discuss in detail in Section \ref{sec:cav_clear}.

For the \textit{no jitter model}, results are similar in most cases to the fiducial model, with the notable exceptions being for CQ Tau and RY Lup. For CQ Tau, a significant small separation signal is observed with the \textit{no jitter model}, in contrast with the insignificant signal, seen with the fiducial model. Similarly for RY Lup, a significant small separation signal is observed with the \textit{no jitter model}, in contrast with the insignificant signal observed with the fiducial model. The same is observed for these two systems with the \textit{UEVA model}. Additionally, with the \textit{UEVA model}, high-mass, small-separation orbits are more likely for AB Aur than lower-mass, larger separation orbits, compared with the fiducial model (and all other models). In contrast, for MWC 758, and to a lesser extent UX Tau A, the \textit{UEVA model} limits the high-mass, small-separation part of parameter space, and favours lower-mass, larger separation orbits. For MWC 758 in particular, the results from the \textit{UEVA model} suggest that the companion that we have detected is {sub-stellar, possibly below the deuterium burning limit} (see Figure \ref{fig:mass-a-UEVA}, in the Appendix). Based on the \textit{UEVA model} posterior, we find that the MWC 758 companion has a {82\%} probability of having a mass less than 75 $M_{Jup}$ and a {22\%} probability of having a mass less than 13 $M_{Jup}$, for orbits that do not cross the inner edge of the cavity ($a(1+e)<a_{cav} = 0.75 \times R_{cav,mm}$).

One potential explanation for the apparent unexplained discrepancy between models with an additional jitter term and those without could be due to multiple massive companions producing a signal that is not captured by our single companion model. {Another possibility, particularly in the case of RY Lup and CQ Tau, is that it is possible that stellar-mass companions on short-period orbits are responsibly for the observed signal in the jitter-free models, and the discrepancy we observe with the jitter models is due to some systematic uncertainty in the astrometry.} Alternatively, another possibility is that we are observing an additional astrometric noise that affects some protoplanetary disk systems, in line with what was seen by \citet{2022RNAAS...6...18F}, who report that the GDR3 re-normalized unit weight error (RUWE) parameter is inflated by disk material. One physical explanation for this unexplained disk signal could be due to inner disk variability \citep[possibly due to variable structure or misalignment, e.g.,][]{2020MNRAS.492..572A,2021A&A...654A..97G,2024ApJ...966..167G,2025AJ....169..318S} producing spatial perturbations to the photocenter. However, this would require significant Gaia G-band flux from inner disks, and it is unclear whether this is common. Notably, RY Lup is one example where some previous investigations predict significant inner disk flux on the $\sim$1 mas scales required to produce an observed effect on Gaia astrometry \citep[e.g.,][]{2020A&A...642A.162G}. If it is confirmed that there is excess noise due to protoplanetary disk inner disks, this will limit our ability to detect low mass companions. However, it may still be possible to model the inner disk contamination in GDR4 data using a geometrical model or a correlated noise model such as a quasi-periodic Gaussian process model. See Section 2.4 from \citet{2025arXiv250110488L}, and Section 5 from \citet{2025arXiv251200157V} for discussions on other possible sources of astrometric noise for young stars.

The constraints on the error inflation parameter, $\sigma_{\mu,jitter}$, from our fiducial model, shown in Figure \ref{fig:jitters}, are in line with the differences we observed between models for CQ Tau and RY Lup. For these two systems, the elevated jitter values indicate a disagreement between the short timescale GDR2, GDR32 and GR3 proper motions and the long timescale HG proper motion, that is compensated for by an increase in $\sigma_{\mu,jitter}$. It is also notable that there is a moderate elevation in the observed 95th percentile upper limit for UX Tau A, {AB Aur} and HD 142527, compared with the rest of the sample. For UX Tau A {in particular}, including the UEVA constraint in the model excludes higher mass and some smaller separation samples, so if the observed slight elevation in the systematic noise is real, it {may be} semi-coherent on the Gaia scan-law timescale of several weeks. We note that the observed error inflation distributions are not correlated with the inclination of the outer disk, which suggests that it is not related directly to asymmetries in the outer disk. 



\subsection{Connection to observed cavities}
\label{sec:cav_clear}

Using the \textit{co-planar jitter model}, in Figure \ref{fig:disk-clear-detections} we present the posterior samples for AB Aur and MWC 758 that are consistent with having cleared the observed disk cavities, under the assumption that there is one companion in the cavity and that companion is the source of the observed astrometric signal. We do not include UX Tau A because we observe no significant constraints on the position of a companion consistent with the astrometry beyond the GDR3 observation epochs (between 2014 and 2017). Additionally, most posterior samples for UX Tau A are not consistent with having cleared the observed cavity due to its likely small semi-major axis (e.g., Figure \ref{fig:mass-a-nogeom-jitter} and Figure \ref{fig:mass-a-UEVA}, in the Appendix). 

To estimate which posterior samples are consistent with the observed cavities, we follow the prescription outlined by \citet{2025A&A...698A.102R}. We take samples with a mass ratio, semi-major axis and eccentricity that are consistent to within 25\% of carving a cavity of size $a_{cav} = 0.75 \times R_{cav,mm}$ (an empirical estimate of the gas cavity size) based on the predictions made with Equations 5 and 6 from \citet{2025A&A...698A.102R}. We also impose that orbits do not cross the gas cavity inner edge ($a(1 +e) < a_{cav}$). 
{We use an eccentricity of zero for the AB Aur cavity and an eccentricity of 0.1 for the MWC 758 cavity \citep{2018ApJ...860..124D}}. We set the mutual inclination term between the disk and companion to zero for mass ratios $>$0.01 for simplicity since we are looking at orbits that are near co-planar with the disk (and because it is neglected in the mass ratio $<$0.01 case), and plot the orbits coloured by mass, along with the predicted positions on 1 March 2026, overplotted on ALMA continuum images of these disks. We present the same for HD 142527, except with all of the posterior samples shown, to highlight our clear detection of the known M dwarf companion in the system, that is known to have an orbit too small to explain the cavity size according to theory \citep{2024A&A...683A...6N}. In Figure \ref{fig:full-pos-preds} in Appendix \ref{sec:full_predictions}, we present the full position predictions of all orbit samples for the same disks shown in Figure \ref{fig:disk-clear-detections}, on 1 June 2015, near the mid-point of the Gaia data, where the position is close to the most constrained, and on 1 March 2026.

Based on the above prescription, we calculate the median plus/minus the 84th/16th percentile of the mass and the semi-major axis for the scenario of a single cavity carving companion from our posteriors. For AB Aur we find $M_b = 76^{+120}_{-56}$ $M_{Jup}$ and $a = 53^{+11}_{-8}$ au; and for MWC 758 we find $M_b = 18^{+9}_{-8}$ $M_{Jup}$ and $a = 17^{+3}_{-3}$ au. For HD 142527, based on the full model posterior, we find $M_b = 186^{+72}_{-50}$ $M_{Jup}$ and $a = 9^{+6}_{-4}$ au, which is {reasonably} consistent with the mass of 273 $\pm$ 13 $M_{Jup}$ and the semi-major axis of 10.8 $\pm$ 0.2 au that we calculate with a new orbit fit to relative astrometric measurements between HD 142527 A and HD 142527 B, along with absolute astrometry data from the HGCA only in Appendix \ref{sec:hd142527_orbit}. {We note that with the \textit{jitter model} (assuming no co-planar constraint), we find a mass of $251^{+160}_{-77}$ $M_{Jup}$ and a semi-major axis of $10^{+2}_{-4}$ au, which is completely consistent with the results we find in Appendix \ref{sec:hd142527_orbit}, suggesting that the slight tension with the \textit{co-planar, jitter model} results could be due to the known misalignment between the orbit of HD 142527 B and the outer disk \citep{2024A&A...682A.101S,2024A&A...683A...6N}.}

With the results shown in Figure \ref{fig:disk-clear-detections} we highlight that by including a reasonable prior on the orbital geometry of a perturbing companion, we are able to place meaningful constraints on the location of a companion that is consistent with the observed astrometric signal and consistent with having carved the disk cavity. These results suggest that targeted observations consisting of multiple science fiber pointings with VLTI/GRAVITY dual-field mode \citep[e.g.,][]{2024A&A...686A.258P} could be made to make the first dynamical mass measurement of an imaged protoplanet (or sub-stellar object) in a protoplanetary disk, or in the case of a non-detection, rule out that the observed astrometric signal is responsible for having cleared the cavity, if sufficient sensitivity is achieved. 

Additionally, our method could be applied to testing the dynamical origin of observed dust features. One notable example would be for suspected dust trapped at Lagrangian points of an unseen protoplanet \citep[e.g.,][]{2022ApJ...937L...1L,2023A&A...675A.172B}. This could be done by calculating the expected planet position based on a dust clump suspected to be located at a Lagrangian point (L4 or L5) of a planet (constrained by previously measured relative astrometry of a dust clump), and using that planet position to calculate the astrometric perturbation on the star to use in our astrometry model. 


\subsection{Comparison to known companions and candidates}
\label{sec:comparisonto_other}

{In this section we compare our fiducial model posterior predictions to known and candidate companions in each system. We emphasize that for our modelling we do not consider direct imaging or radial velocity constraints due to concerns about dust extinction and stellar activity. Therefore, our constraints should be considered upper limits---particularly at very small and very large separations. More comprehensive constraints on the presence of companions can be made on a source by source basis by carefully combining our astrometry results with limits on small separation companions from radial velocity measurements, and limits on large separation companions from direct imaging \citep[e.g.,][]{2022A&A...665A..73M,2025arXiv250110488L}.}

\begin{figure}
\includegraphics[width=1\linewidth]{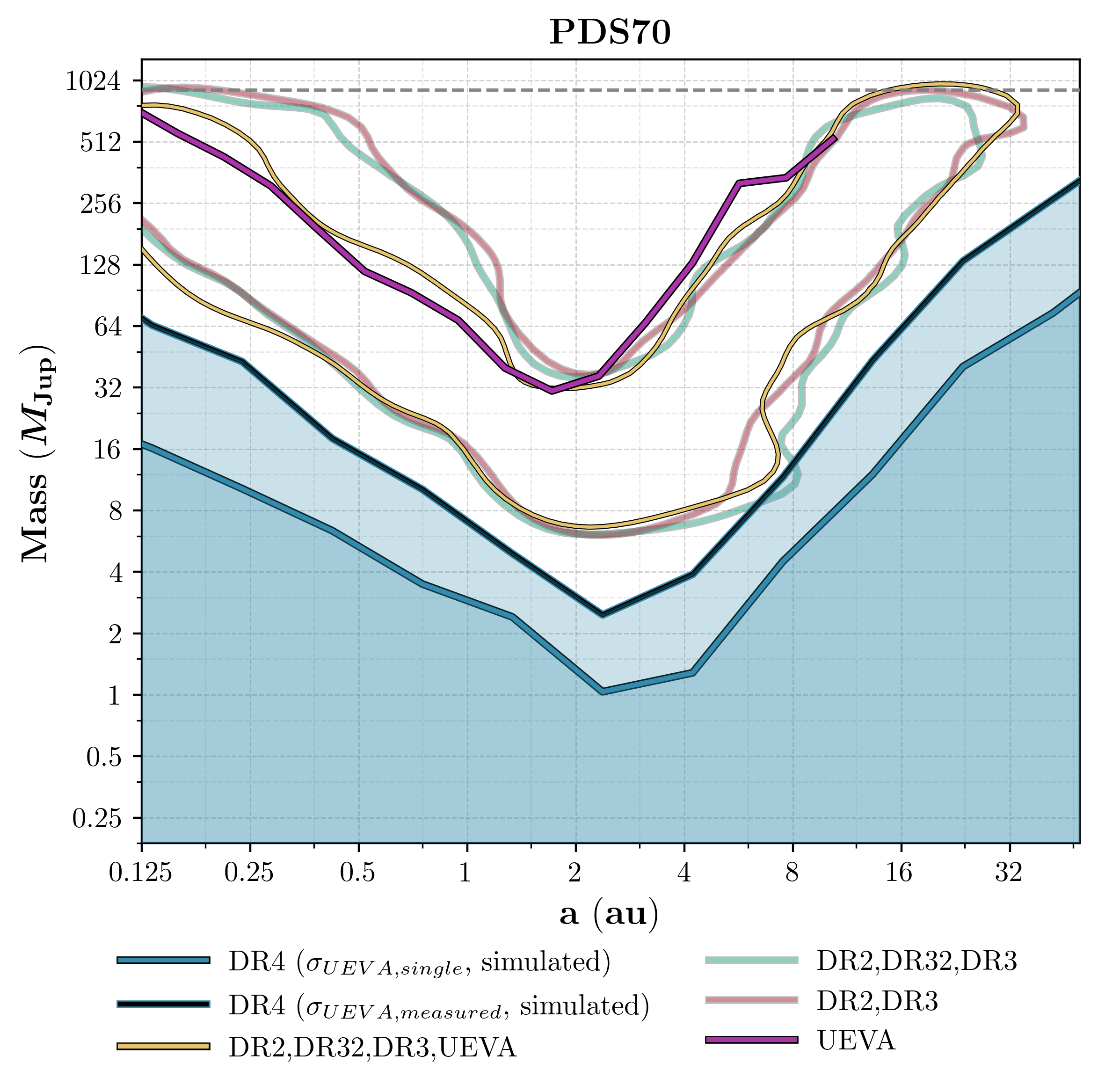}
\caption{GDR2 and GDR3 constraints and GDR4 predicted constraints for PDS 70. The yellow, and the (semi-transparent) teal and red contours denote the 1$\sigma$ posterior density constraints for models that use the DR2,DR32,DR3,UEVA, DR2,D32,DR3, and the DR2,DR3, subsets of the G23H catalog. The magenta curve shows the 95th percentile upper limit from the UEVA constraint, showing no significant signal. The blue filled curve denotes the 95th percentile upper limit calculated from the GDR4 simulation using the single star noise model. The black filled curve shows the same for the measured UEVA noise model. The dashed grey line denotes the stellar mass of 0.88 $M_{\mathrm{\odot}}$.
}
\label{fig:pds70}
\end{figure}

\begin{figure}
\includegraphics[width=1\linewidth]{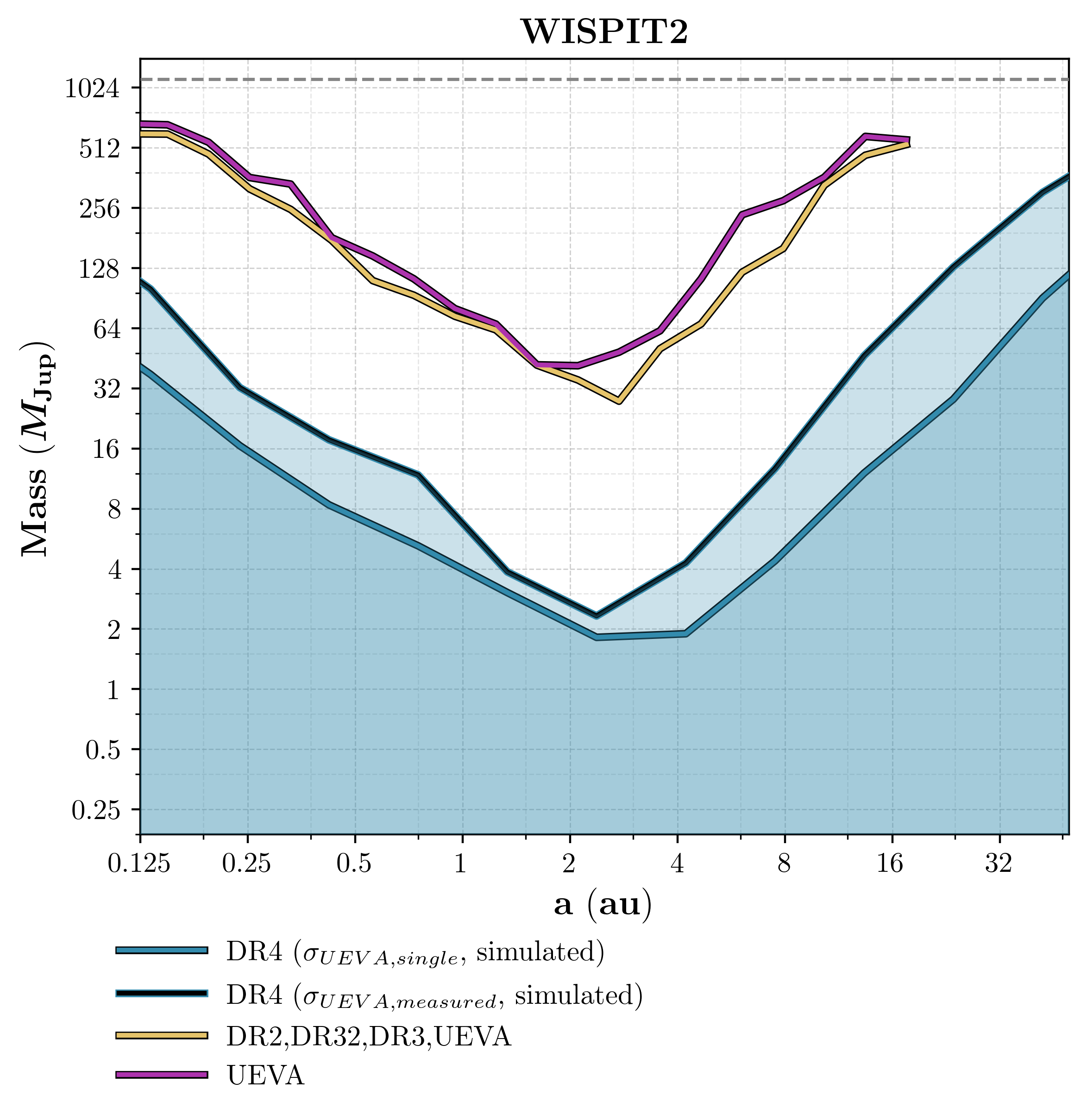}
\caption{
GDR2 and GDR3 constraints and GDR4 predicted constraints for WISPIT 2. The yellow and the magenta curves show the 95th percentile upper limit from the DR2,DR32,DR3,UEVA, and the UEVA constraints, respectively, showing no significant signal. The blue filled curve denotes the 95th percentile upper limit calculated from the GDR4 simulation using the single star noise model. The black filled curve shows the same for the measured UEVA noise model. The dashed grey line denotes the stellar mass of 1.08 $M_{\mathrm{\odot}}$. Note that there is no significant constraint from the GDR2, GDR32 or GDR3 proper motions, so the individual model results are omitted and the DR2,DR32,DR3,UEVA model constraint is due solely to the UEVA. 
}
\label{fig:wispit2}
\end{figure}

\textit{HD 142527:} For all four of the models we explore we find constraints that are {mostly} consistent with the measured $\sim$11 au semi-major axis \citep{2024A&A...682A.101S,2024A&A...683A...6N}, and $273 \pm 13$ $M_{Jup}$ mass of HD 142527 B (see Appendix \ref{sec:hd142527_orbit}). Additionally, the predicted positions we find with the \textit{co-planar, jitter model} are {reasonably} consistent with previous direct imaging and interferometric detections, despite the measured mutual inclination of $\sim$10-20 degrees between the orbital plane and the outer disk \citep{2024A&A...682A.101S,2024A&A...683A...6N}.

\textit{MWC 758:} 
The point-like source identified by \citet{2018A&A...611A..74R} is not consistent with the observed astrometric signal, assuming it is an object that is near co-planar with the disk. At the time of their observations (2015-2016), our \textit{co-planar, jitter model} predicts a massive companion near to the opposite side of the star for samples similar to the candidate identified by \citet{2018A&A...611A..74R}, with a separation of $\sim$110 mas ($\sim$20 au). Notably, the scenario of a $\sim$10 $M_{Jup}$ object on an eccentric orbit in the mm dust cavity that was shown to be consistent with driving the observed near-infrared spiral arms by \citet{2020MNRAS.498..639C} is reasonably consistent with the astrometric signal that we have identified with all four of our models. However, the predicted position of the companion scenario simulated by \citet{2020MNRAS.498..639C} is at a position roughly south, south-west of the primary at a separation of $\sim$200 mas, which is not consistent with what our models predict. 
The directly imaged planet candidate MWC 758 c is at too large a separation to be constrained by the astrometry \citep{2023NatAs...7.1208W}. We note that \citet{2025arXiv251200157V} do report a significant (4.4$\sigma$) GDR2-GDR3 proper motion anomaly for MWC 758, however, it is not clear whether the signal that the authors report is consistent with our detection given that there is some tension between our mass$-$semi-major axis constraints.

\textit{AB Aurigae:} 
AB Aur b \citep{2022NatAs...6..751C,2022ApJ...934L..13Z,2025AJ....169..258B,2025ApJ...990L..42C} {lies at too large an angular separation to be consistent with the co-planar predictions for a sub-stellar object} (see Figures \ref{fig:disk-clear-detections} and \ref{fig:full-pos-preds}). Additionally, if AB Aur b is a planetary mass-object as is suggested by its photometry \citep{2022NatAs...6..751C}, then we have no significant constraint on its astrometric signature. {Similarly, the co-planar model predictions are inconsistent with the position of the companion candidate identified by \citet{2026AJ....171..250K}, at a separation of 65 au and a position angle of 147$^{\circ}$ in 2024.}
In contrast, the predicted position from the \textit{co-planar jitter model} is reasonably consistent with the {``twist feature" and \textit{feature f1}} seen by \citet{2020A&A...637L...5B} (see Figure \ref{fig:full-pos-preds}). However, this companion would not be consistent with clearing the observed {disk cavity based on the prescription from \citet{2025A&A...698A.102R} (see Figure \ref{fig:disk-clear-detections})}. {The astrometry is also consistent with the scenario envisioned by \cite{2020MNRAS.496.2362P}, of a stellar-mass companion on a highly inclined, eccentric orbit, which is predicted to be at a separation of $\sim$180 mas towards the east-southeast of the primary star.} Finally, \cite{2024MNRAS.534.2904C} suggest that the disk features observed with ALMA may be consistent with a {stellar or sub-stellar} mass companion, which we do find is consistent with the observed astrometric signal.

\textit{UX Tau A:} The $\sim$400 au projected separation in the plane of the sky between UX Tau A and UX Tau C (B is at an even larger separation from A) \citep{2020A&A...639L...1M} is too large for a perturbation from C to explain any of the observed astrometric signal.

\textit{CQ Tau:} The {upper limits we observe with the fiducial model} could be consistent with an inner, eccentric companion similar to the scenario suggested by \citet{2022MNRAS.515.6109H}.

\textit{HD 97048:} 
The kinematically detected planet by \citet{2019NatAs...3.1109P}, at a separation of $\sim$130 au, is at too large a separation for the astrometry to place any constraint on its mass.

\textit{HD 100546:} Our constraints rule out stellar mass companions with a semi-major axis greater than roughly {2 au}, which is inconsistent with a stellar mass (mass ratio of 0.3) companion on an 8.4 au semi-major axis, 0.6 eccentricity orbit, as proposed by \cite{2022ApJ...936L...4N}, or the stellar mass (mass ratio of 0.1) companion with a semi-major axis of 10 au and an eccentricity of 0.4 proposed by \cite{2024MNRAS.534.2904C}. These constraints are also in tension with a significant fraction of the parameter space of the photometrically estimated roughly 25-50 $M_{Jup}$ companion candidate identified by \cite{2025AJ....169..152B}, or would imply that the object has a mass closer to 10 $M_{Jup}$. The candidate identified by \cite{2015ApJ...807...64Q} is too far out for our method to provide meaningful constraints. {Additionally, our upper limits would imply a mass of no more than $\sim$10 $M_{\mathrm{Jup}}$ for the candidates identified by \cite{2019ApJ...883...37B} and \cite{2015ApJ...814L..27C}}. 

\textit{RY Lup:} Our astrometry models that consider systematic jitter provide very weak constraints on any companions in the RY Lup disk. The considered models place no constraints on the possibility of a $\sim$2$M_{Jup}$ companion at 190 au considered by \citet{2018A&A...614A..88L}.

\textit{HD 100453:} All of the models considered in this work show a clear non-detection of any significant astrometric signal for HD 100453. This is in contrast to the apparent 3.8$\sigma$ {detection of a possibly} planetary mass companion claimed by \citet{2025arXiv251200157V}, based on the uncalibrated GDR2 and GDR3 proper motions and the GDR3 RUWE parameter. Our models rule out the majority of the mass -- semi-major axis parameter space consistent with the signal that the authors report. HD 100453 is an important example of how the assumptions made for the method used by \citet{2025arXiv251200157V} could lead to false positives.

\textit{HD 34282:} Our models are not able to rule out the 50 $M_{Jup}$ companion at $\sim$30 au, suggested by \citet{2017A&A...607A..55V}. We do rule out similar mass companions on circular orbits between $\sim$5 and $\sim$25 au, extending the (local extinction limited) near-infrared sparse aperture masking limits presented by \citet{2025ApJ...993..178V} to lower masses at smaller separations.

\textit{TW Hya:} Our models constrain the mass of any companion in the disk of TW Hya to at most several Jupiter masses, between $\sim$3-10 au. Our constraints are not sensitive enough to rule out the expected several tens of Earth mass ($\sim$0.1 $M_{Jup}$) planets that could be responsible for carving the observed dust gap at $\sim$6 au \citep[e.g.,][]{2017ApJ...837..132V}.


\textit{PDS 70:} 
We do see an apparently significant signal in the GDR2, GDR32 and GDR3 proper motions of PDS 70, shown in Figure \ref{fig:pds70}. 
However, due to the large orbital separations of PDS 70 b and c we find no meaningful dynamical mass constraint on the mass of PDS 70 b \citep[semi-major axis of $\sim$20 au,][]{2025A&A...698A..19T}, or PDS 70 c  \citep[semi-major axis of $\sim$32 au,][]{2025A&A...698A..19T}. For the candidate PDS 70 d, which has an orbit with a semi-major axis of $\sim$13 au  \citep{2025MNRAS.539.1613H}, the constraint we find from the joint GDR2, GDR32, GDR3, UEVA model, places a 1$\sigma$ lower limit on its mass to be $\sim$70 $M_{Jup}$. This high mass for the candidate d is in significant tension with the mass constraints found by \citet{2025A&A...698A..19T}. Thus, we conclude that the astrometric signal we observe is not due to the planet candidate d. Finally, \citet{2025AJ....169..137B} do detect excess emission at an SNR of $\sim$4, with an unconstrained separation in JWST/NIRISS F480M aperture masking interferometry data. It is possible that (at least some of) the astrometric signal we observe for PDS 70 is related to the object identified by \citet{2025AJ....169..137B}, however we leave investigating this to a future work.

\textit{WISPIT 2:} The protoplanet WISPIT 2 b, at a deprojected separation of $\sim$54 au, and the inner companion candidate CC1, at a deprojected separation of $\sim$15 au, \citep{2025ApJ...990L...8V,2025ApJ...990L...9C} are at too large separations for any meaningful dynamical mass constraints to be made with the Gaia data alone. We present our dynamical mass constraints for WISPIT 2 in Figure \ref{fig:wispit2}, along with the expected limits from GDR4.

\subsection{Gaia DR4 Predictions}
To estimate the expected sensitivity limits for transition disks with GDR4 epoch astrometry, we look at two well known protoplanet hosts PDS 70 and WISPIT 2, which are both too faint to be in the Hipparcos catalog. 
In Figures \ref{fig:pds70} and \ref{fig:wispit2} we present the Gaia only constraints (discussed above) and predictions for GDR4 under two different noise assumptions.

We simulate and model GDR4 data for WISPIT 2 and PDS 70 using \texttt{Octofitter v8}. To estimate individual astrometric measurements we first download the GDR4 scan law using  the Gaia Observation Forecast Tool (GOST)\footnote{https://gaia.esac.esa.int/gost/index.jsp}. Next, to simulate the individual CCD measurements we draw a random number of CCD measurements per scan as $n_{ccds} \sim Poisson(N)$, where $N$ is the average number of scans per window (which is roughly 9), and take  $max(1, min(n_{ccds}, 9))$ as the number of CCD measurements for each scan. We assume the time spread of each scan window is 40 seconds and assume that the scan angle changes linearly by 0.1 degrees over each window. We repeat this procedure 10000 times and take a result with the median number of measurements. 

To estimate the noise on each individual measurement we use the square root of the UEVA values calculated from the G23H catalog. We use the UEVA to calculate noise estimates for GDR4 because it is an unbiased estimator of the typical individual measurement error estimated from Gaia DR3 data \citep{2025A&A...702A..76K}. The two noise values we consider are for a single star with the same magnitude, colour and sky position as PDS 70 and WISPIT 2 ($\sigma_{UEVA,single} = \sqrt{UEVA_{single}}$), and directly from the Gaia DR3 RUWE values ($\sigma_{UEVA,measured} = \sqrt{UEVA_{measured}}$), assuming that any additional astrometric noise seen for either star is not due to perturbations from any massive inner companions and is effectively normally distributed noise that is intrinsic to each system (the known companions are at separations too large to explain any observed signal). 

We simulate the GDR4 data by assuming a single star with proper motion and parallax given by the GDR3 values, and a stellar mass of  $0.88 \pm 0.09$ $M_{\odot}$ for PDS 70 \citep{2021AJ....161..148W} and $1.08 \pm 0.11$ $M_{\odot}$ for WISPIT 2 \citep{2025ApJ...990L...8V}, assuming 10\% uncertainties on the stellar masses. We note that this procedure returns 595 individual CCD measurements and $\sigma_{UEVA,single} = \sqrt{UEVA_{single}} = 0.138$ mas for Gaia BH3, which has prerelease GDR4 data that has been made publicly available \citep{2024A&A...686L...2G}. The prelease data consists of 599 individual CCD measurements (excluding outliers), with a mean (median) statistical uncertainty of 0.142 mas (0.135 mas), which is reasonably consistent with our estimate. We do not consider the $\sigma_{UEVA,measured}$ value because the $\sqrt{UEVA_{measured}}$ value is of course dominated by the 33 $M_{\odot}$ black hole that is perturbing the star \citep{2024A&A...686L...2G}. Note that we do not consider missed scans, which means that the distribution (in time) of our simulated measurements is not identical, but the effect this has on the estimated limits should be small. 

To improve sampling for companions at large separations due to the degeneracy between a large separation companion and the proper motion of the system, we analytically marginalize over the position offset and proper motion of the system. We do this by subtracting a weighted least-squares fit to the residuals before calculating the likelihood.


Due to the limited time baseline that the GDR4 data covers, only companions on the left hand side of roughly the minimum mass shown in both Figures \ref{fig:pds70} and \ref{fig:wispit2} will have reasonably constrained orbits. To the right hand side of the minimum, only a curvature will be able to be measured, and thus there will be a degeneracy between the companion mass and separation, with a lower mass companion on a shorter orbit producing an identical curvature to a higher mass companion on a longer orbit. Our predictions extend to larger separations than those in \citet{2026AJ....171...18L} by also considering detections of acceleration only, in addition to closed orbits. 
We note that {direct imaging limits and} constraints due to the presence of dust/gas features for protoplanetary disk targets will significantly limit the parameter space of companions responsible for any observed astrometric signal, similar to the {approaches taken by \citet{2022A&A...665A..73M} and} \citet{2025NatAs...9.1176R}.

From the GDR4 simulations, we see that---even in the pessimistic case (using $\sigma_{UEVA,measured}$) of all of the observed astrometric noise being due to something intrinsic about protoplanetary disks---epoch astrometry from GDR4 data will allow for the presence of Jupiter-like planets to be constrained in protoplanetary disk systems.

\section{Conclusions} \label{sec:conclusions}
In this work, we present dynamical mass constraints for companions in 11 transition disk systems using the G23H absolute astrometry catalog. We clearly detect the known low mass stellar companion HD 142527 B at very high significance (odds ratio $>$ 1000 in favour of the companion model for our fiducial model) as a proof of concept. With absolute astrometry alone, we find an orbit and mass of HD 142527 B that is consistent with what we find for its orbit by combining literature relative astrometry and the HGCA. From this new orbit fit, we find a dynamical mass of $273 \pm 13$ $M_{Jup}$, which is the most precise dynamical mass estimate of HD 142527 B to date. 

{Using the G23H catalog, we newly detect a likely sub-stellar companion of MWC 758, at a high significance (odds ratio $>$ 80). We additionally find moderate evidence of companions to AB Aur and UX Tau A (odds ratios $>$ 3). For HD 100453, HD 100546, HD 97048, HD 34282 and TW Hya we find no significant evidence for astrometric signals produced by companions (odds ratios $<$ 3), and place strong constraints on the parameters of a companion that could be responsible for the observed cavities in these systems.} We note that our clear non-detection of a significant astrometric signal for HD 100453 (with any of our models) is in contradiction with the 3.8$\sigma$ detection claimed by \citet{2025arXiv251200157V}. We suspect that this result may be due to known systematic biases in the method used by \citet{2025arXiv251200157V} that could be significant for low-mass companions (see Section \ref{sec:introduction} for details). 
For CQ Tau and RY Lup, we find no significant evidence for astrometric signals in two out of the four models that we consider, {both of which include a jitter term to account for potential systematics}. For both CQ Tau and RY Lup, with the other two models we find a short timescale signal that is in tension with the long timescale signal between Hipparcos and Gaia measurements. This indicates that either there are multiple massive companions in the system producing a signal not captured by our single companion model, {there are truly short-period stellar companions in these systems and there is some level of systematic uncertainty in the astrometry,} or there is an unexplained source of excess astrometric noise, seen for some protoplanetary disk systems.

We also run detailed simulations of Gaia DR4 data for PDS 70 and WISPIT 2. We show that DR4 will not have the sensitivity to significantly constrain the dynamical masses of the known planets in either system \citep{2018A&A...617A..44K,2019NatAs...3..749H,2025ApJ...990L...8V,2025ApJ...990L...9C}; however, we do show that DR4 data will be sensitive to planets not too dissimilar from Jupiter ($a \sim 1-5$ au, $M \gtrsim 2-5$ $M_{Jup}$), even for protoplanetary disk systems with distances greater than 100 pc. Our results demonstrate the strong possibility of being able to determine the comprehensive demographics of a new class of protoplanets in protoplanetary disk systems with the release of Gaia DR4.

\begin{acknowledgments}

This work has made use of data from the European Space Agency (ESA) mission Gaia (\href{https://www.cosmos.esa.int/gaia}{https://www.cosmos.esa.int/gaia}), processed by the Gaia Data Processing and Analysis Consortium (DPAC)\footnote{https://www.cosmos.esa.int/web/gaia/dpac/consortium}.
This work made use of the Gaia Observation Forecast Tool (GOST).

This research used the facilities of the Canadian Astronomy Data Centre operated by the National Research Council of Canada with the support of the Canadian Space Agency.

This research was enabled in part by support provided by the Digital Research Alliance of Canada (alliancecan.ca).

This paper makes use of the following ALMA data: ADS/JAO.ALMA\#2021.1.00690.S, 
2015.1.01137.S, 2017.1.00492.S.

We thank the referee for their careful reviewing of this paper. D.J.\ is supported by NRC Canada and by an NSERC Discovery Grant. 
J.S.\ acknowledges financial support from the Natural Sciences and Engineering Research Council of Canada (NSERC) through the Canada Graduate Scholarships Doctoral (CGS D) program, and from the Heising–Simons Foundation through the 51 Pegasi b Fellowship (grant \#2025-5885).

\end{acknowledgments}

\appendix

\section{Full position predictions}
\label{sec:full_predictions}
\begin{figure*}
\centering
\includegraphics[width=0.95\textwidth]{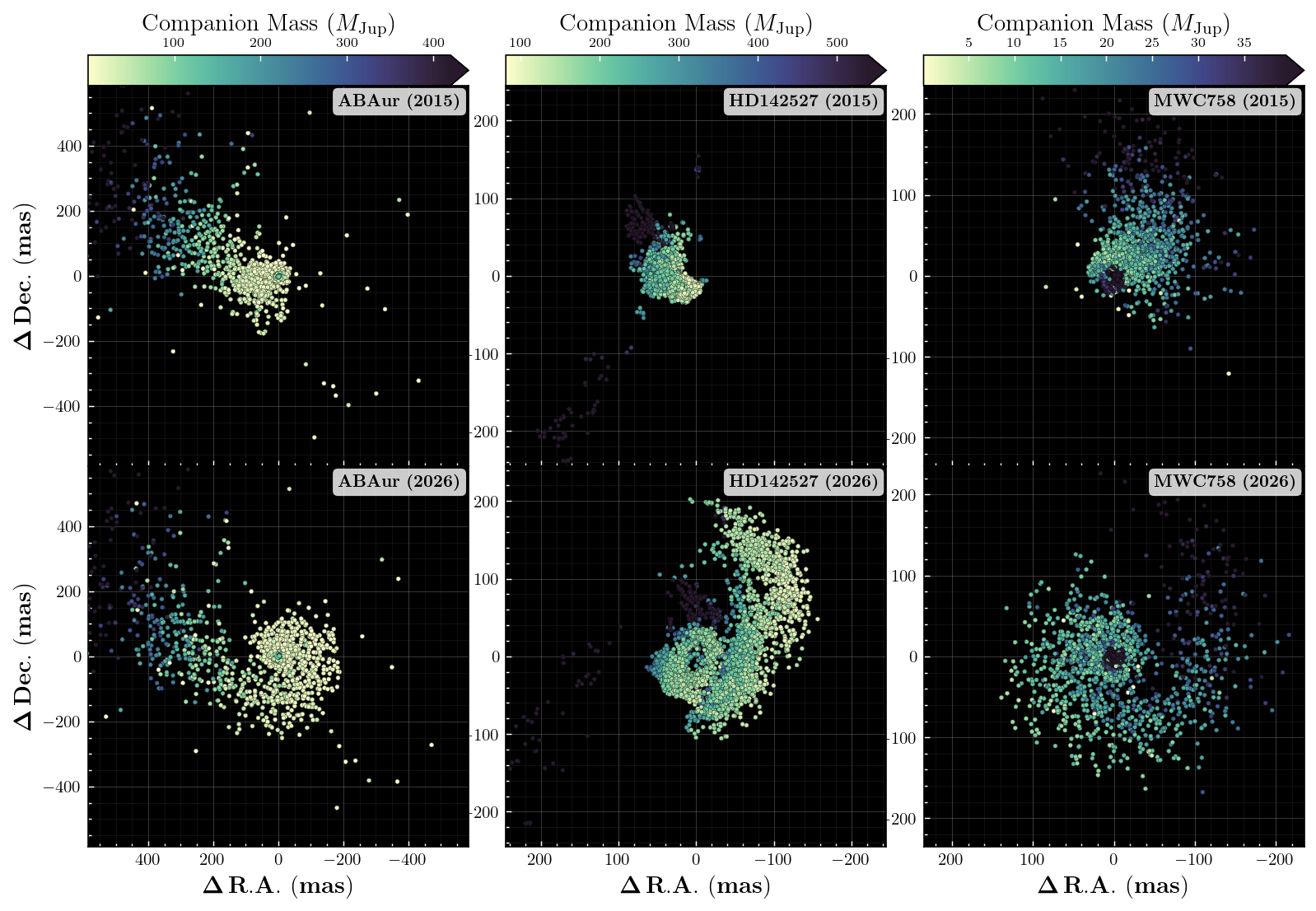}
\caption{Posterior samples from the \textit{co-planar, jitter model}, coloured by mass. We show the predicted positions on 1 June 2015, near the mid-point of the Gaia DR3 data, and on 1 March 2026. We show all samples, but clip the colour scale to the 5th and 95th percentile of the mass distribution for orbits with a semi-major axis greater than 2 au for visual clarity.
}
\label{fig:full-pos-preds}
\end{figure*}

In Figure \ref{fig:full-pos-preds}, we present the predicted positions from the full posterior of the \textit{co-planar, jitter model} on 1 June 2015 and 1 March 2026.

\section{Marginal posteriors}
\label{sec:mass-a-post}

\begin{figure*}
\centering
\includegraphics[width=0.82\textwidth]{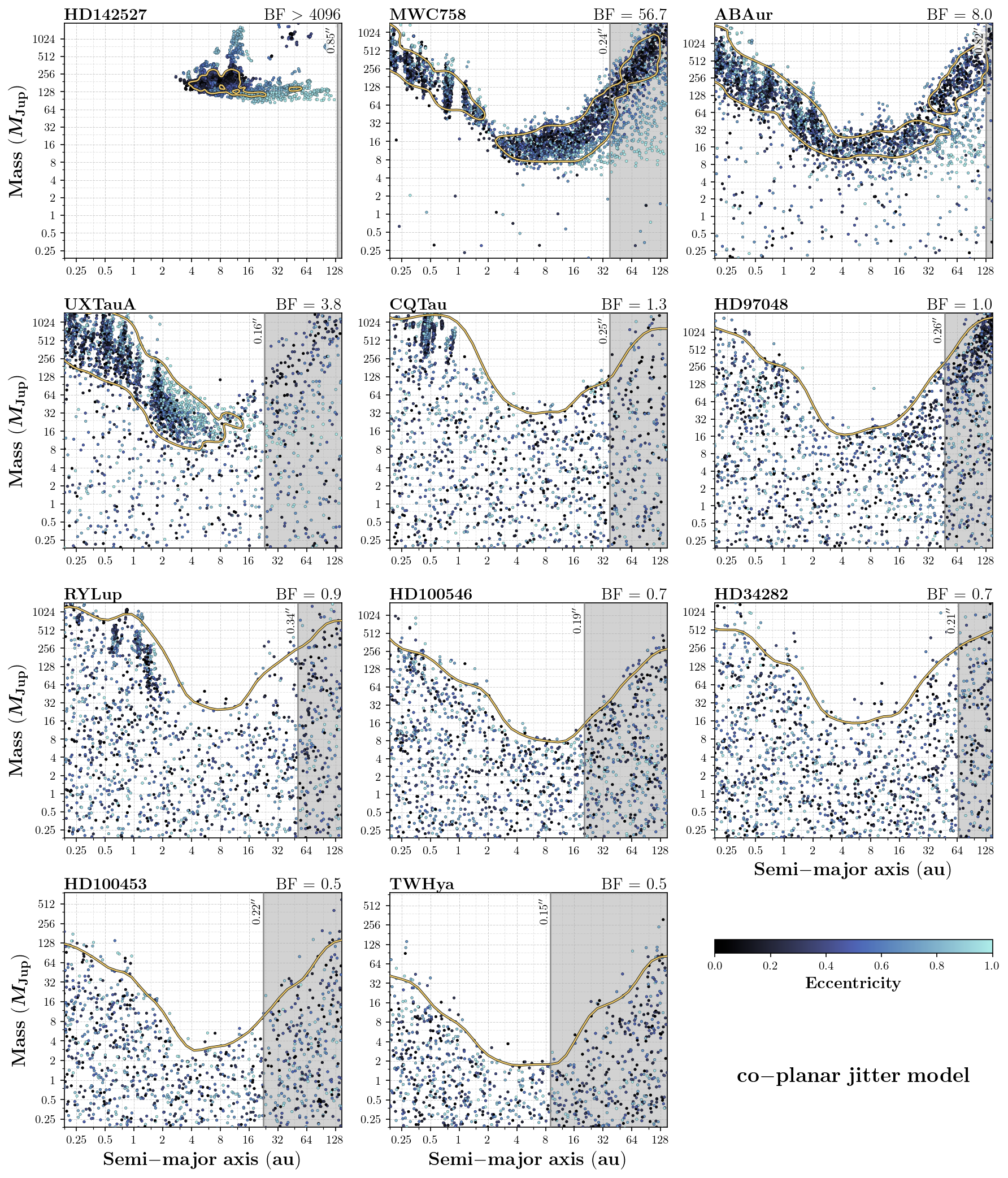}
\caption{Companion mass and semi-major axis marginal posterior for the \textit{co-planar, jitter model}. For stars with BF $>$ 3, the gold contour represents the 1$\sigma$ posterior density. For stars with BF $<$ 3, the gold curve represents the 95th percentile upper limit. The shaded region shows 0.75$\times R_{cav,mm}$, an empirical estimate of the gas cavity size \citep{2025A&A...698A.102R}.}
\label{fig:mass-a-geom-jitter}
\end{figure*}

\begin{figure*}
\centering
\includegraphics[width=0.82\textwidth]{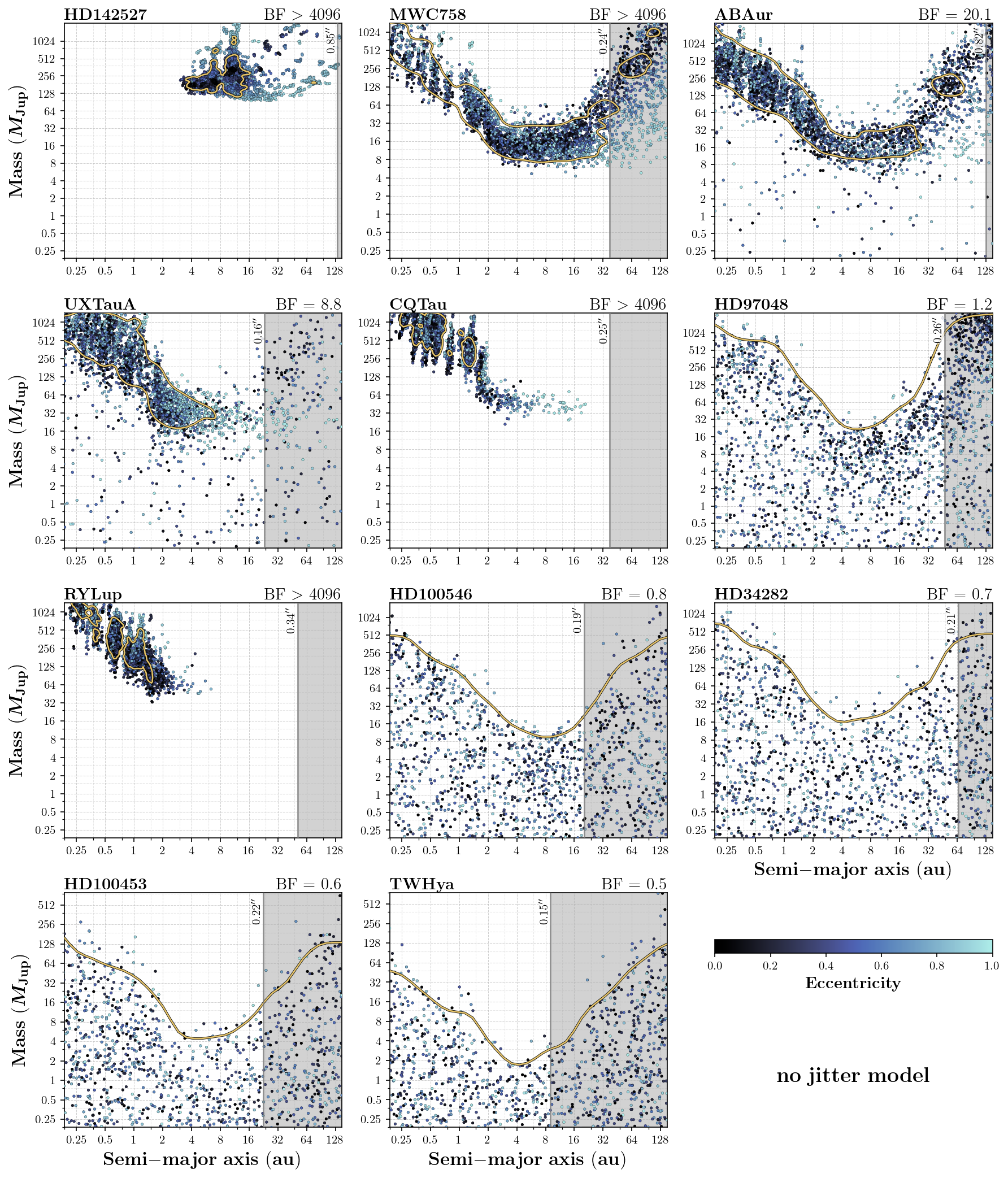}
\caption{Identical to Figure \ref{fig:mass-a-geom-jitter}, for the \textit{no jitter model}.}
\label{fig:mass-a-nojitter}
\end{figure*}

\begin{figure*}
\centering
\includegraphics[width=0.82\textwidth]{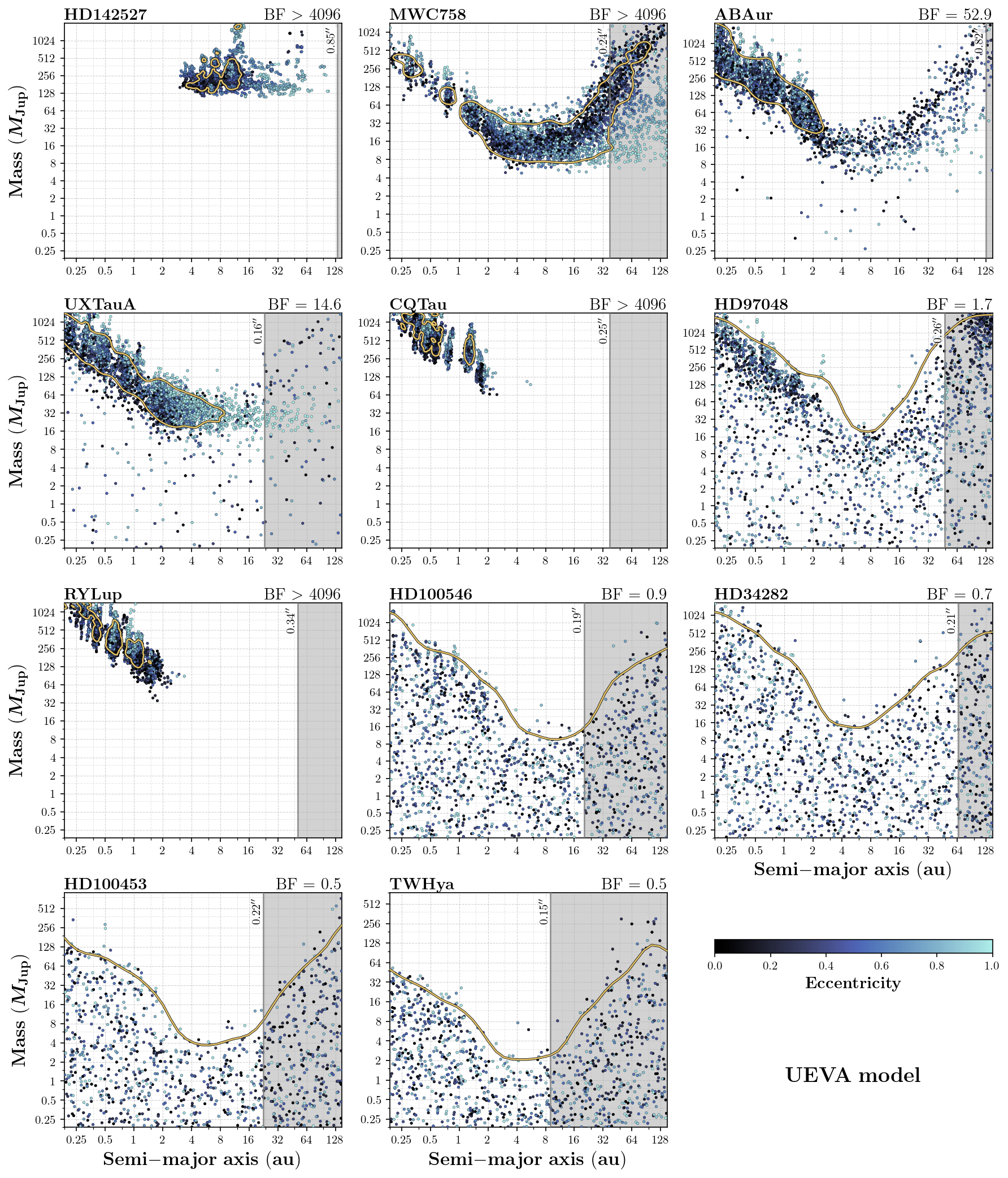}
\caption{Identical to Figure \ref{fig:mass-a-geom-jitter}, for the \textit{UEVA model}.}
\label{fig:mass-a-UEVA}
\end{figure*}

In Figures \ref{fig:mass-a-geom-jitter}, \ref{fig:mass-a-nojitter} and \ref{fig:mass-a-UEVA}, we present the mass versus semi-major axis marginal posteriors for the \textit{co-planar, jitter model}, the \textit{no jitter model} and the \textit{UEVA model}, respectively.



\section{The dynamical mass of HD 142527 B}
\label{sec:hd142527_orbit}

To calculate the dynamical mass of HD 142527 B (and HD 142527 A), we run an orbit fit to literature relative astrometry consisting of the measurements reported in Table 2 from \citet{2024A&A...683A...6N}, which includes measurements from \citet{2012ApJ...753L..38B}, \citet{2014ApJ...791L..37R}, \citet{2016A&A...590A..90L}, \citet{2018A&A...617A..37C}, \citet{2019A&A...622A..96C},
\citet{2019A&A...622A.156C} and \citet{2022AJ....164...29B}. Additionally, we include the 2018 and 2019 relative astrometry measurements in Table 1 from \citet{2024A&A...682A.101S}. We also include the HGCA \citep{2021ApJS..254...42B} proper motions to constrain the individual masses. We use the same priors on orbital elements as are displayed in Table \ref{tab:priors}, with the exception being that we use a uniform prior on the primary mass between 1 $M_{\odot}$ and 3 $M_{\odot}$. We also do not consider the flux of HD 142527 B. 

\begin{figure*}
\centering
\includegraphics[width=0.5\textwidth]{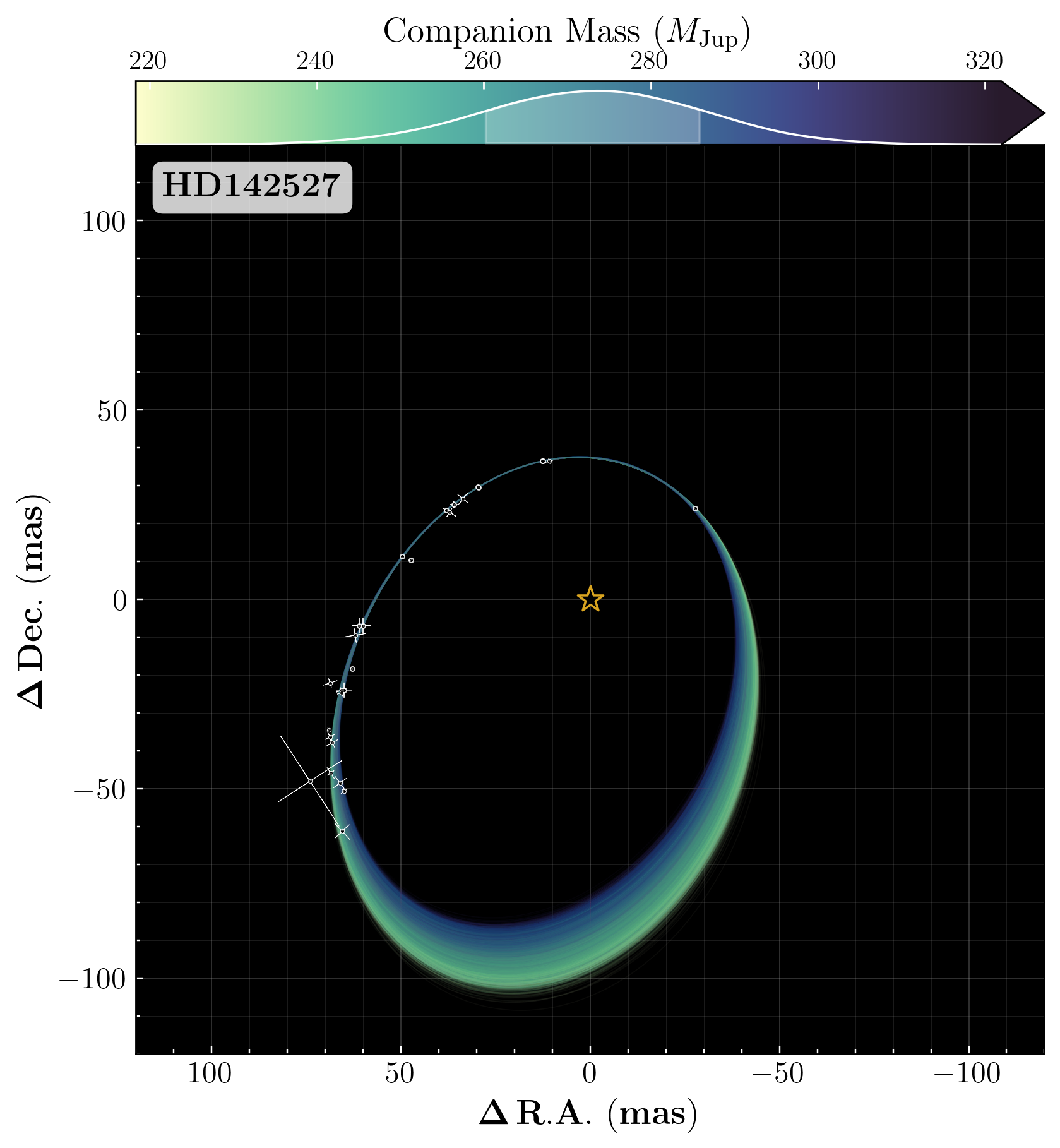}
\caption{Posterior samples coloured by mass from our joint relative astrometry and HGCA orbit model for HD 142527 B. The position of HD 142527 A is shown by the gold star.}
\label{fig:hd142527orbit}
\end{figure*}

Posterior samples coloured by mass, along with the relative astrometry are shown in Figure \ref{fig:hd142527orbit}. We find a dynamical mass of  $2.16 \pm 0.04$ $M_{\odot}$ for HD 142527 A, and $273 \pm 13$ $M_{Jup}$ for HD 142527 B. Additionally, we constrain the orbital eccentricity to $0.46 \pm 0.01$, the semi-major axis to $10.8 \pm 0.2$ au, the inclination to $148.9 \pm 0.6$ degrees, and the longitude of the ascending node to $160 \pm 2$ degrees, assuming that the plane of the orbit is near co-planar to the outer disk. The orbital constraints we find are consistent with the results found by \citet{2024A&A...683A...6N}.

We note that we only use the HGCA absolute astrometry (along with the relative astrometry) for this analysis so as to calculate a (partially) independent mass measurement to compare with what we find using only absolute astrometry from the G23H catalog. We leave the determination of a dynamical mass of HD 142527 B with the G23H catalog to a future work.

\section{The HGCA Model}
\label{sec:hgca_model}

\begin{figure*}
\centering
\includegraphics[width=0.82\textwidth]{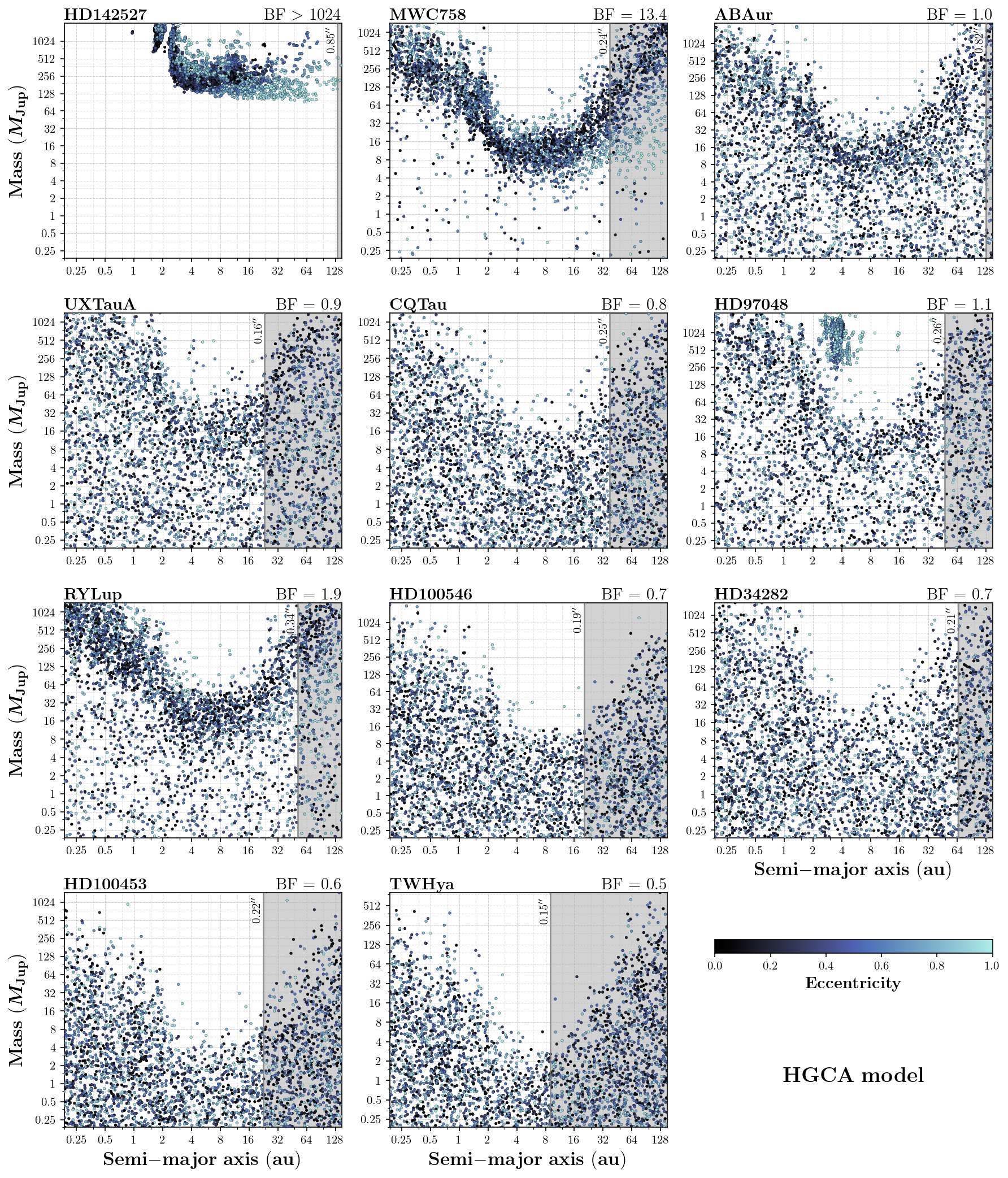}
\caption{{Companion mass and semi-major axis marginal posterior for the \textit{HGCA model}. The shaded region shows 0.75$\times R_{cav,mm}$, an empirical estimate of the gas cavity size \citep{2025A&A...698A.102R}.}}
\label{fig:mass-a-HGCA}
\end{figure*}

{In this appendix, we present results for the \textit{HGCA model}, that uses only the HGCA \citep{2021ApJS..254...42B} and the Hipparcos IAD \citep{2007A&A...474..653V,2020AJ....159...71N}. 
In Figure \ref{fig:mass-a-HGCA}, we present the mass versus semi-major axis marginal posteriors for the \textit{HGCA model}.} 

\bibliography{sample701}{}
\bibliographystyle{aasjournalv7}



\end{document}